\documentclass[aps,prl,preprint,superscriptaddress,floatfix]{revtex4}
\usepackage{times}
\usepackage{graphicx}
\usepackage{epsfig}
\usepackage{amsmath}
\usepackage{amssymb}
\usepackage{natbib}
\usepackage{color}

\begin{document}

\title{Structural Complexity and Correlated Disorder in Materials Chemistry}

\author{Andrew~L.~Goodwin}
\email{andrew.goodwin@chem.ox.ac.uk}
\affiliation{Department of Chemistry, University of Oxford, Inorganic Chemistry Laboratory, South Parks Road, Oxford OX1 3QR, U.K.}
\date{\today}

\date{\today}



\maketitle

\section{Introduction}

Complexity is a measure of information content \cite{Kolmogorov_1968}. Crystalline materials are not complex systems because their structures can be represented tersely using the language of crystallography \cite{Cartwright_2012}. Disordered materials are also structurally simple if the disorder present is random: such systems can be described efficiently through statistical mechanics \cite{Ziman_1979}. True complexity emerges when structures are neither perfectly crystalline nor randomly disordered---a middle ground once named ``organised complexity'' \cite{Weaver_1948}. In current parlance, in our field, we use the term ``correlated disorder'' for this same regime, emphasising the presence and importance of non-random patterns \cite{Keen_2015}.

Correlated disorder now seems more widespread in materials chemistry than originally thought \cite{Simonov_2020}. Our increasing sensitivity to its existence comes primarily from advances in X-ray detectors, which now allow routine measurement of the weak diffuse scattering that is its signature \cite{Welberry_2015}. Sometimes this diffuse scattering---and hence the underlying disorder---has a straightforward origin and is not relevant to properties of particular interest. At other times, however, disorder is crucial for material function. Modern electrode materials \cite{Liu_2024}, thermoelectrics \cite{Roth_2021}, and metamaterials \cite{Gartside_2022} offer many examples, and we will come to discuss a few others drawn from our own work below. Such cases give a strong imperative for materials chemists to understand disorder/property relationships, and to develop chemical control over the presence and nature of correlated disorder in different materials \cite{Simonov_2020}.

An imperative is all well and good, but why should such control even be possible? After all, it is challenge enough to control the structures of ordered materials \cite{Pitcher_2015}. The cause for optimism comes from the realisation that complexity can arise from very simple ingredients \cite{Crutchfield_2012}. In the context of correlated disorder in materials chemistry, the challenge becomes one of identifying the relevant ingredients---chemical, electronic, geometric---responsible for particular kinds of correlated disorder. Control over disorder then follows from control over those components \cite{Simonov_2020}.

\section{Chemical control over correlated disorder}

\begin{figure}
\begin{center}
\includegraphics[width=\textwidth]{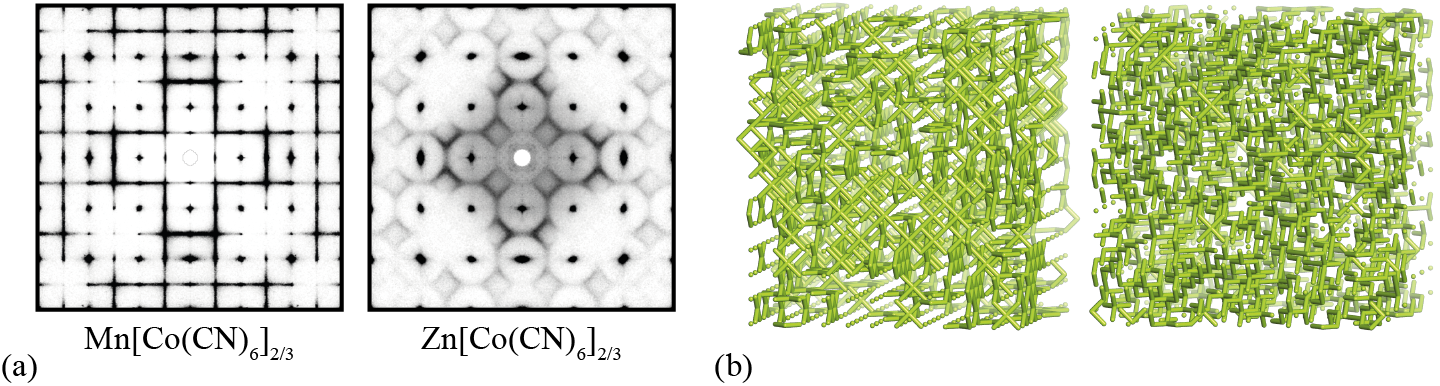}
\end{center}
\caption{\label{fig1} Correlated vacancy disorder and micoropore networks in PBAs. (a) PBAs of different composition or synthesised under different conditions show contrasting diffuse scattering patterns in X-ray diffraction, indicative of distinct correlations within their disordered micropore networks. (b) Two representative micropore networks with contrasting connectivity and tortuosity. These topological differences lead to different mass transport characteristics.}
\end{figure}

We first realised the potential for such control when studying vacancy disorder in Prussian blue analogues (PBAs) \cite{Simonov_2020b}. Vacancies are important in these materials, as they connect to form the micropore networks involved in many aspects of PBA function: e.g. catalysis \cite{Peeters_2012}, gas storage \cite{Kaye_2005}, guest-dependent magnetism \cite{Ohkoshi_2010}, and charge storage \cite{Lu_2012}. Using the X-ray scattering methods outlined above, we were able to show that the rate at which PBA samples are prepared affects how uniformly vacancies are distributed, with quickly-grown crystals having increasingly random arrangements. At the same time, PBA composition affects geometric patterns in these arrangements with different transition-metals favouring or disfavouring the preservation of local inversion symmetry---as expected from their corresponding $d$-electron counts (Fig.~\ref{fig1}(a)). Since the resulting micropore networks have different characteristics---such as connectivity, anisotropy, or tortuosity---we now know how chemistry and synthetic strategy can be combined to target PBAs with specific mass-transport properties (Fig.~\ref{fig1}(b)). We think this control will be particularly important when optimising the performance of PBA-based cathodes \cite{Wu_2016}.

In the field of computational soft-matter, a variety of mechanisms for control over complexity in colloidal systems are being uncovered \cite{Damasceno_2012}; one of particular interest involves interactions favouring two separate distances that compete to drive geometric frustration \cite{Dshemuchadse_2021}. Nature has no simple structural solutions to resolve these competing requirements, leading to the suppression of crystallisation and the emergence of complexity. We were able to show that this paradigm is unexpectedly relevant to amorphous calcium carbonate (ACC), an important precursor in biomineralization processes \cite{Addadi_2003,Nicholas_2024}. In ACC, carbonates mediate the interaction between Ca$^{2+}$ ions to favour two distinct Ca$\ldots$Ca separations: one at $\sim$4\,\AA\ and the other at $\sim$6\,\AA, corresponding to cases where Ca$^{2+}$ ions are bridged by a single oxygen atom or by different oxygens of the same CO$_3^{2-}$ ion, respectively \cite{Goodwin_2010}. The ratio of $\sim$1.5 is amongst those that drive complexity. So it is a simple consequence of the radii and coordination modes of Ca$^{2+}$ and CO$_3^{2-}$ that calcium carbonate resists crystallisation---unlike related salts such as calcium fluoride or copper sulfate that crystallise easily. This mapping to soft-matter theory also suggests how varying geometric ingredients, such as ionic radii or molecular geometry, might tune the balance between order and disorder in ACC analogues.

Our third example draws on the interplay of order and disorder found in tilings first described by Sébastien Truchet over 300 years ago \cite{Truchet_1704}. Truchet’s insight was to show how complexity could emerge when high-symmetry tiles are decorated to lower their symmetry. As chemists, we are familiar both with tilings---used to describe intermetallics, quasicrystals, and zeolites---and with symmetry-lowering mechanisms such as Jahn--Teller distortions, chirality, or polarisation. We recently showed how combining these ideas allows synthesis of atomic-scale analogues of Truchet tilings. We used the chemistry of metal–organic frameworks (MOFs) to do so, with the very simple strategy of lowering the point symmetry of the organic linkers from which they are assembled \cite{Guillerm_2019}. In this way, the periodic structure of MOF-5 \cite{Li_1999} evolves into the aperiodic Truchet network TRUMOF-1 as linear 1,4-benzenedicarboxylate linkers are replaced by their bent 1,3-substituted congener \cite{Meekel_2023}. The resulting tiling---combining cubes and trigonal bipyramids---is decorated by a symmetry-lowering connectivity, yielding a non-repeating but chemically sensible architecture. This aperiodicity enhances mechanical properties \cite{Meekel_2024}, but more importantly establishes a Truchet-inspired design strategy for complexity that extends beyond MOFs.

\begin{figure}
\begin{center}
\includegraphics[width=\textwidth]{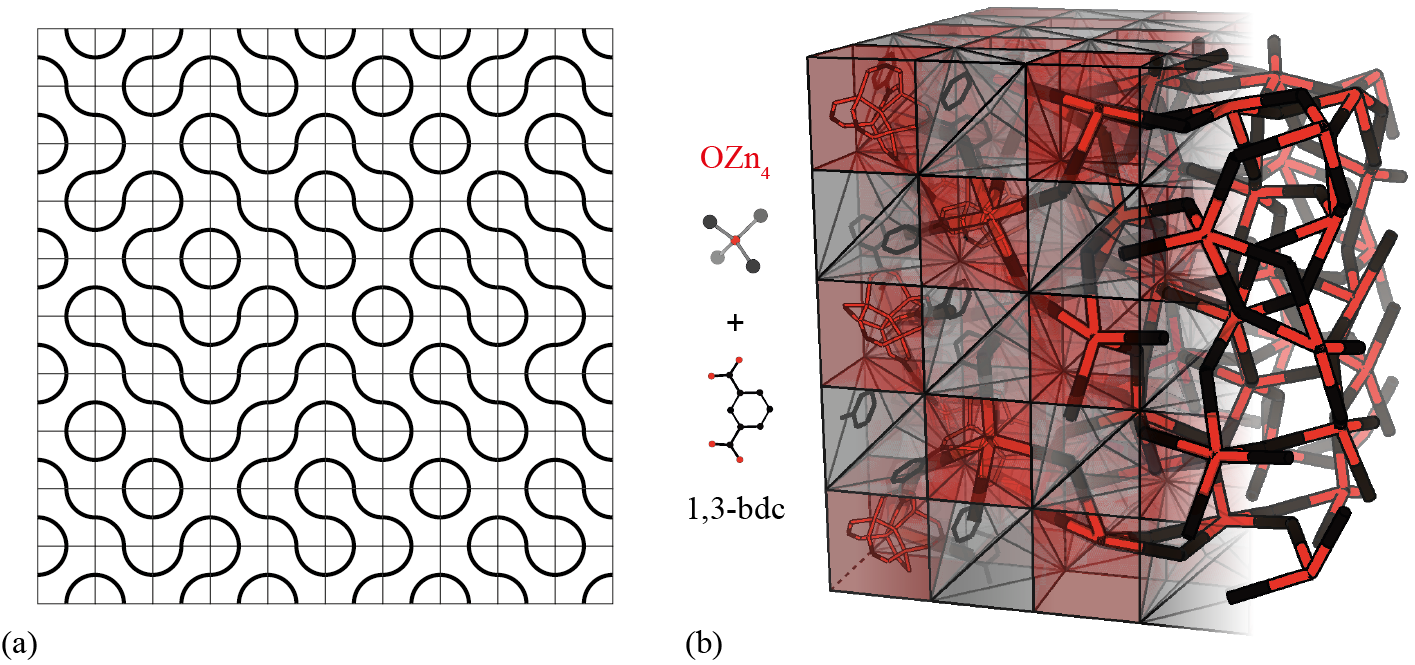}
\end{center}
\caption{\label{fig2} Truchet tilings and their atomic-scale analogue in TRUMOF-1. (a) Truchet tilings are assembled from high-symmetry tiles, decorated to lower their symmetry. The decorations form aperiodic patterns that store visual information. (b) The Truchet-like structure of TRUMOF-1 arises when OZn$_4$ clusters are connected by 1,3-bdc linkers. The positions of nodes and linkers correspond to a periodic tiling of space, but the network traces a disordered labyrinth.}
\end{figure}

\section{Outlook and future challenges}

The central role of geometry in each of these examples has an important consequence: namely, the phenomenology of correlated disorder is likely to generalise across quite different chemistries. Already it is known, for example, that the rules governing vacancy distributions in PBAs also determine substitutional disorder in Heusler thermoelectrics \cite{Roth_2021} and lithium-ion pathways in disordered rocksalt cathodes \cite{Schmidt_2024}. Likewise the Truchet-tile paradigm connects naturally to long-studied models in statistical mechanics \cite{Wannier_1950} and frustrated magnetism \cite{Lieb_1967}, which have been applied to systems as diverse as supramolecular assemblies \cite{Blunt_2008}, ferroelectrics \cite{Grindlay_1959}, and negative thermal expansion materials \cite{Ramirez_2000}. So, although the field of correlated disorder lacks the formalism of crystallography, the emergence of a framework for distinguishing different kinds of correlated disorder is becoming clear. We believe this framework will help the subject, and its role in materials chemistry, to mature rapidly over the coming years.

Looking forward, we envisage an increasing shift from characterising and understanding correlated disorder in functional materials towards exploiting it to generate functionalities not accessible to ordered states. An obvious direction here is to capitalise on the information content inherent to complex systems. In the same spirit that Truchet tilings are related to QR codes and barcodes \cite{Smith_1987}, so might one view individual TRUMOF-1 crystals as storing vast amounts of information in their particular three-dimensional connectivity. We have highlighted previously the potential advantage of the redundancy of distributed information storage in such networks \cite{Reynolds_2021}, which connects this direction with the development of error-correcting codes \cite{Pretzel_1992}. For this avenue to bear fruit we will need transformative methods for encoding information in disordered systems, and similarly transformative methods for reading the information contained. These are clear challenges for the community.

In our own group, we are currently focusing on a new concept of ``responsive disorder'', where we are seeking to understand how external stimuli might drive meaningful changes in the type of correlated disorder present. The motivation here is that many functional responses of ordered materials are linked to phase transitions, for which again there is a strong language in terms of symmetry breaking processes \cite{Landau_1980}. Now that we know many different disordered states are possible, shouldn’t we look for phenomena associated with transitions between them? Such transitions may help explain “hidden order” anomalies reported in some magnetocalorics \cite{Schiffer_1995} and superconductors \cite{Kung_2015}, or connect to the topological order parameters invoked in models of liquid structure \cite{Neophytou_2022}. Given that nonlinearity and memory effects are more prevalent in complex systems \cite{Crutchfield_2012}, we envisage the eventual development of ``adaptive disorder'', in which material response evolves with history. We are beginning to explore this idea directly, using correlated disorder as a platform for physical reservoir computing \cite{Gartside_2022}. Whatever the precise course the field comes to take, harnessing the information content of correlated disorder may yet prove to be one of the most powerful design strategies in materials chemistry.

\section{Acknowledgements}

The author gratefully acknowledges funding from the European Research Council (Advanced Grant 788144), the UKRI through Frontier Research Grant EP/Z534031/1, and Royal Society for funding through the Faraday Discovery Fellowships Fund, provided by DSIT.

\end{document}